\documentclass[runningheads]{llncs}
\usepackage[T1]{fontenc}
\usepackage{amsmath}
\usepackage{amssymb}
\usepackage{algorithm}
\usepackage{algorithmic}
\usepackage{float}
\usepackage{graphicx}
\usepackage{mathtools}
\usepackage{pgffor}
\usepackage[numbers, sort]{natbib}
\usepackage{fancyhdr}
\usepackage[title,titletoc]{appendix}
\usepackage{xcolor}

\usepackage{soul}
\reversemarginpar

\newcommand{\be}{\begin{equation}}
\newcommand{\ee}{\end{equation}}

\makeatother
\graphicspath{{figures/}}

\begin{document}
\title{Wasserstein Image Local Analysis: Histogram of Orientations, Smoothing and Edge Detection}
%
%
\author{Jiening Zhu\inst{1} \and
Harini Veeraraghavan\inst{2} \and
Larry Norton\inst{2} \and
Joseph O. Deasy\inst{2} \and
Allen Tannenbaum\inst{1}}
\authorrunning{J. Zhu et al.}
%
\institute{Stony Brook University, NY, United States
\email{\{jiening.zhu,allen.tannenbaum\}@stonybrook.edu} \and
Memorial Sloan Kettering Cancer Center, NY, United States \\
\email{\{veerarah,nortonl,deasyj\}@mskcc.org}}
\maketitle              
\begin{abstract}
The Histogram of Oriented Gradient (HOG) is a widely used image feature, which describes local image directionality based on numerical differentiation. Due to its ill-posed nature, small noise in the image may lead to large errors. Thus conventional HOG may fail to produce meaningful directionality results in the presence of noise, which is  common in medical radiographic imaging. We approach the directionality problem from a novel perspective by the use of the optimal transport map of a local image patch to a uni-color patch of its mean. Concretely, we decompose the transport map into sub-work costs each transporting in different directions. To test our approach, we evaluated the ability of the optimal transport to quantify tumor heterogeneity from MRI brain images of patients diagnosed with glioblastoma multiforme available from the TCIA. By considering the entropy difference of the extracted local directionality within tumor regions, we found that patients with higher entropy in their images, had statistically significant worse overall survival (p $=0.008$), which indicates that tumors that have images exhibiting flows in many directions may be more malignant, perhaps reflecting high tumor histologic grade, a reflection of histologic disorganization.
We also explored the possibility of solving classical image processing problems such as smoothing and edge detection via optimal transport. By looking for a 2-color patch with minimum transport distance to a local patch, we derive a nonlinear shock filter, which preserves edges. Moreover, we found that the color difference of the computed 2-color patch indicates whether there is a large change in color, i.e., an edge in the given patch. In summary, we expand the usefulness of optimal transport as an image local analysis tool, to extract robust measures of imaging tumor heterogeneity for outcomes prediction as well as image pre-processing required in several image analysis applications. Because of its robust nature, we find it offers several advantages over the classical approaches.

\end{abstract}
\section{Introduction}
Image feature extraction is an classical topic ever since image data was first digitalized. Over the past several decades, multiple different image feature describers have been proposed. Steerable filters are synthesized filters of arbitrary orientations from linear combinations of basis filters \cite{93808}. Scale-Invariant Feature Transform (SIFT) and Histogram of Orientated Gradient (HOG) are two feature describers based on histogram of gradients \cite{Lowe2004,QIQQA-UKDEU}. Image feature describers are widely used for object detection, object matching, and network pretraining. A similar concept is used in the fields of medical imaging and radiomics. Co-occurrence of Local Anisotropic Gradient Orientations (CoLlAGe) uses a similar histogram of gradient method to capture subtle differences between benign and pathologic phenotypes on anatomic imaging \cite{prasanna_co-occurrence_2016}.

Wasserstein distance is a powerful metric for distributions, known to be robust compared to other metrics (or divergences). It is defined naturally by the optimal cost of transporting one distribution to another, which was first motivated by the civil engineering problem of relocating a pile of soil to an excavation site by Gaspard Monge in 1781 \cite{villani1,villani2,Evans1999,Rachev1998}. A relaxation was proposed by the Russian mathematician Leonid Kantorovich \cite{LK} in 1942, and so the optimal transport problem is many times called the {\em Monge-Kantorovich problem}. It gives a distance of the space of probability distribtions, called the {\em Wasserstein distance}. In this note, we will only use the $L^1$ version, namely, the $W_1$ metric. 

Optimal mass transport methods are widely used in signal processing, machine learning, computer vision, meteorology, statistical physics, quantum mechanics, and network theory \cite{Arjovsky2017,CarMaa14,Haker2004,MitMie16,rr,Statement2020}. In the medical field, a number of works incorporate the Wasserstein distance and mass transport for various purposes. Some examples include \cite{Zhu2021.06.17.448878,9266088} who use the distance to study multiomics networks for cancer subtype clustering, \cite{romt} that used regularized optimal transport to visualize fluid flows in the glymphatic system, as well as \cite{aapm}, which employed the unbalanced Wasserstein distance to identify high-risk normal tissue regions associated with worse mortality from spillover radiation in radiation treatments.

In this work, we make use of the robust property of the Wasserstein distance for image local analysis in order to extract directional information from the optimal transport map instead of gradient. The optimal transport based computation increases robustness to noise, making it preferable for a variety of medical image analysis applications. Furthermore, we show the applicability of this methodology for other classical image processing tasks, namely, smoothing and edge detection.

In the following sections, we sketch some of the background on optimal transport and and then detail our proposed methodology. Quantitative comparison and illustrative examples of our method are included in the experiments and results section. We conclude this note with some discussion about further applications to pathology data. 

\section{Background and Methods}

\subsection{Background: optimal transport}

The original Monge's formulation of optimal transport may be given a modern expression in terms of measure theory as follows \cite{villani1, villani2}:
\begin{equation} \label{eq:opt1}
\mathcal{W}_M(\mu_0,\mu_1)=\inf_T\{\int_{\mathcal{S}}c(x,T(x))d\mu_0(x)\ |\ T_{\#}\mu_0=\mu_1\},
\end{equation}
where $\mathcal{S}$ denotes a subdomain of $\mathbb{R}^n$, $\mu_0,\mu_1$ are two measures on $\mathcal{S}$, $T: \mathcal{S}\rightarrow\mathcal{S}$ is the transport map, and $c(\cdot,\cdot):\mathcal{S}\times\mathcal{S}\rightarrow \mathbb{R}_{\geq 0} $ is a convex cost function. In the present note, $c(\cdot,\cdot)$ will always be taken to be the {\em \bf distance function}. Here $T_\#$ denotes the push-forward of $T$ ($\mu_1(E)=\mu_0(T^{-1}(E)), \forall E\subset\mathcal{S}$).

Kantorovich relaxed the model by replacing transport maps $T$ in (\ref{eq:opt1}) by couplings $\pi$:
\begin{equation}\label{kantorovich}
  \mathcal{W}_K(\mu_0,\mu_1)=\inf_{\pi\in\Pi(\mu_0,\mu_1)}\int_{\mathcal{S}}c(x,y)\pi(dx,dy),
\end{equation}
where $\Pi(\mu_0,\mu_1)$ denotes the set of all the couplings between $\mu_0$ and $\mu_1$ ( measures on $\mathcal{S}\times \mathcal{S}$ whose two marginals are $\mu_0$ and $\mu_1$: $\pi(\cdot\times \mathcal{S})=\mu_0(\cdot), \pi(\mathcal{S}\times \cdot)=\mu_1(\cdot)$). Despite the relaxation, one may show that Kantorovich and Monge formulations are equivalent in a number of cases under certain continuity constraints; see \cite{villani1, villani2} and the references therein.

There are two major benefits of using Kantorovich form. First, it expands the possible transport maps. Mass can be split up, which Monge's form cannot handle. This relaxation guarantees a solution while Monge's form admits no solution in some cases. Second, in terms of discrete probability densities, Kantorovich form can be written as a standard linear programming problem, which may be solved very efficiently. The discrete form of Kantorovich optimal transport can be expressed as follows:
\begin{subequations}\label{K}
  \begin{align}
  d_K(\rho_0,\rho_1)=\min_{\Pi\in\mathbb{R}^{M\times N}} &\sum_{i=1}^{M}\sum_{j=1}^{N}C(i,j)\cdot \Pi(i,j)\label{eq:k_ob}\\
  \mbox{subject to:}\ &\Pi \vec{1}_N=\rho_0,\label{eq:k_cs1}\\
&\Pi^T \vec{1}_M=\rho_1,\label{eq:k_cs2}
\end{align}
\end{subequations}
where $\rho_0\in\mathbb{R}^M_{+}, \rho_1\in\mathbb{R}^N_{+}$ are density of $M/N$ desecrate locations of initial/target distribution and their total mass needs to be preserved as a prerequisite ($\sum_{i=1}^{M}\rho_0(i)=\sum_{j=1}^{N}\rho_1(j)$). $C$ is a point to point cost matrix and $\vec{1}_{N}(\vec{1}_{M})$ is an all-1 vector of length $N(M)$.


$\Pi$ specifies all the possible couplings with given marginals. The mass initially located at location $i\in \{1,2,...,M\}$ in initial distribution may be transported to any one or more of the locations $j\in\{1,2,...,N\}$ in target distribution. $\Pi(i,j)$ specifies that amount of sub-transport mass and $w(i,j)=C(i,j)\cdot\Pi(i,j)$ is the workload of that sub-transport route. The overall work is the sum of all the work from all the sub-transport routes. So the sub-transport work matrix $W=C.*\Pi$ reveals much of local information, which is very useful.

\subsection{Transport to 1-color patch: directionality}
Usually, the Wasserstein distance is computed globally, analyzing the difference between a pair of images. Here, also a key contribution, we employ optimal transport locally as a nonlinear filter, analyzing the image local features. 

For each local $n\times n$ image patch, we take $\rho_0\in \mathbb{R}_{+}^{n^2}$ to be the image intensity on that patch. An uni-color patch with the mean value of $\rho_0$ as intensity of all pixels on that patch is used as $\rho_1\in \mathbb{R}_{+}^{n^2}$. Further, the cost matrix $C\in \mathbb{R}^{n^2}\times\mathbb{R}^{n^2}$ is taken as the standard pairwise distance matrix. With these definitions, the sub-transport work matrix $W\in \mathbb{R}^{n^2}\times\mathbb{R}^{n^2}$ can be solved via (\ref{K}).

\begin{figure}[H]
  \centering
  \includegraphics[width=\linewidth]{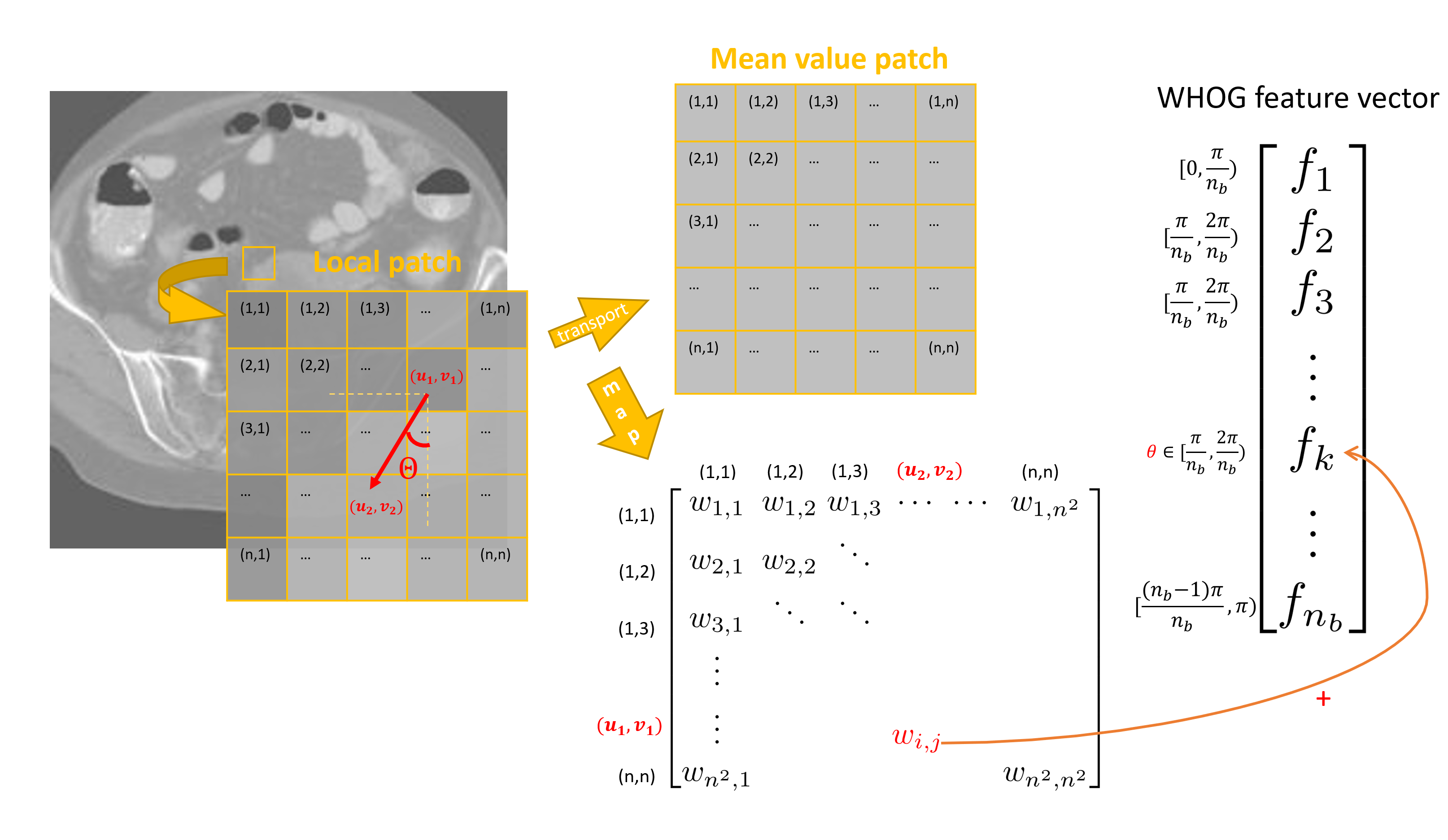}
  \caption{An illustration of the proposed WHOG method.}\label{fig1}
\end{figure}

Local image directions are extracted from the matrix $W$ by regrouping its entries into bins (Fig. \ref{fig1}). Each entry of $W$ has a corresponding orientation given by the transport starting point and end point locations. For the $(i,j)$ entry of $W$ ($i\neq j$), $i$'th location in the density vector corresponding to a coordinate $(u_1,v_1)$ in original $n\times n$ grid and $j$'th location in the density vector corresponding to another coordinate $(u_2,v_2)$ in original $n\times n$ grid. The {\em directionality} of that sub-transport route is given as:
\begin{equation}\label{theta}
  \theta=\arctan(\frac{v_1-v_2}{u_1-u_2}).
\end{equation}

We evenly divide $[0,\pi]$ into $n_b$ bins and add up all the $w_{ij}$ values in their corresponding bins. So each bin contains all the sub-work of the optimal transport in its corresponding direction. The $n_b$-vector is a representation of local directionality. The sum of $n_b$ values in the feature vector coincides with the sum of all $w_{ij}$, which is the Wasserstein distance. On the other hand, we decompose the distance into components in the $n_b$ directions. This gives a similar feature vector that characterizes local directionality as in HOG but more robust. We call our method Wasserstein HOG or WHOG, since its direction extraction is based on the Wasserstein distance. We performed detailed comparisons in the experiments and results section below.

\subsection{Transport to 2-color patches: smoothing, edge detection}
In addition to treating the transport between each local patch to a pure color patch, we also explore the transport from a local patch to 2-color patches. We seek the 2-color patch that has the minimum Wassertein distance to a local patch. The optimization problem may be expressed as follows:

\begin{subequations}\label{K_2}
  \begin{align}
  \min_{\Pi\in\mathbb{R}^{n^2\times n^2},z\in \{0,1\}^{n^2},a,b\in \mathbb{R}} &\sum_{i=1}^{n^2}\sum_{j=1}^{n^2}C(i,j)\cdot \Pi(i,j),\\
  {\mbox subject to:}\ &\Pi \vec{1}_{n^2}=\rho_0,\\
&\Pi^T \vec{1}_{n^2}=a*z+b*(1-z),
\end{align}
\end{subequations}
where $\rho_0$ is the intensity vector of a local patch. $\rho_1=a*z+b*(1-z)$ gives a 2-color patch that is closest to the original patch in term of the Wasserstein distance. Since $\rho_1$ has only two colors, it is smoothed with edge preservation at the same time. Moreover, the intensity difference $|a-b|$ of that patch measures whether there is a large change in color, i.e., an edge in the given patch.

In terms of the sub-transport work matrix $W$. We look at its projection onto the initial and target patches, which gives a an estimate pf pixelwise noise. Ideally, most of the mass stays at the same place and the work is just zero. Pixels that are considered to be noise are those that need extra work to move additional mass away or get mass from other pixels.

\section{Experiments and Results}
\subsection{Compare with conventional HOG}

As a quick review of conventional HOG: The Sobel filters
\begin{equation*}
    \begin{bmatrix}
      1 & 0 & -1
    \end{bmatrix},
    \begin{bmatrix}
      1 \\
      0 \\
      -1
    \end{bmatrix}
  \end{equation*}
  give x- and y-derivatives ($g_x, g_y$) for each pixel. The directionality of that pixel is given by the direction of the gradient vector:
  \begin{equation}\label{theta_hog}
      \theta = \arctan(\frac{g_y}{g_x}).
  \end{equation}
Then all the pixels are put into corresponding bins in terms of $\theta$. The sums of gradient vectors' norms $g=\sqrt{g_x^2+g_y^2}$ of all bins are combined together forming a feature vector of that patch. Feature vectors of neighboring patches are normalized together to alleviate artifacts caused by light variations.

\begin{figure}[H]
  \centering
  \includegraphics[width=0.48\linewidth]{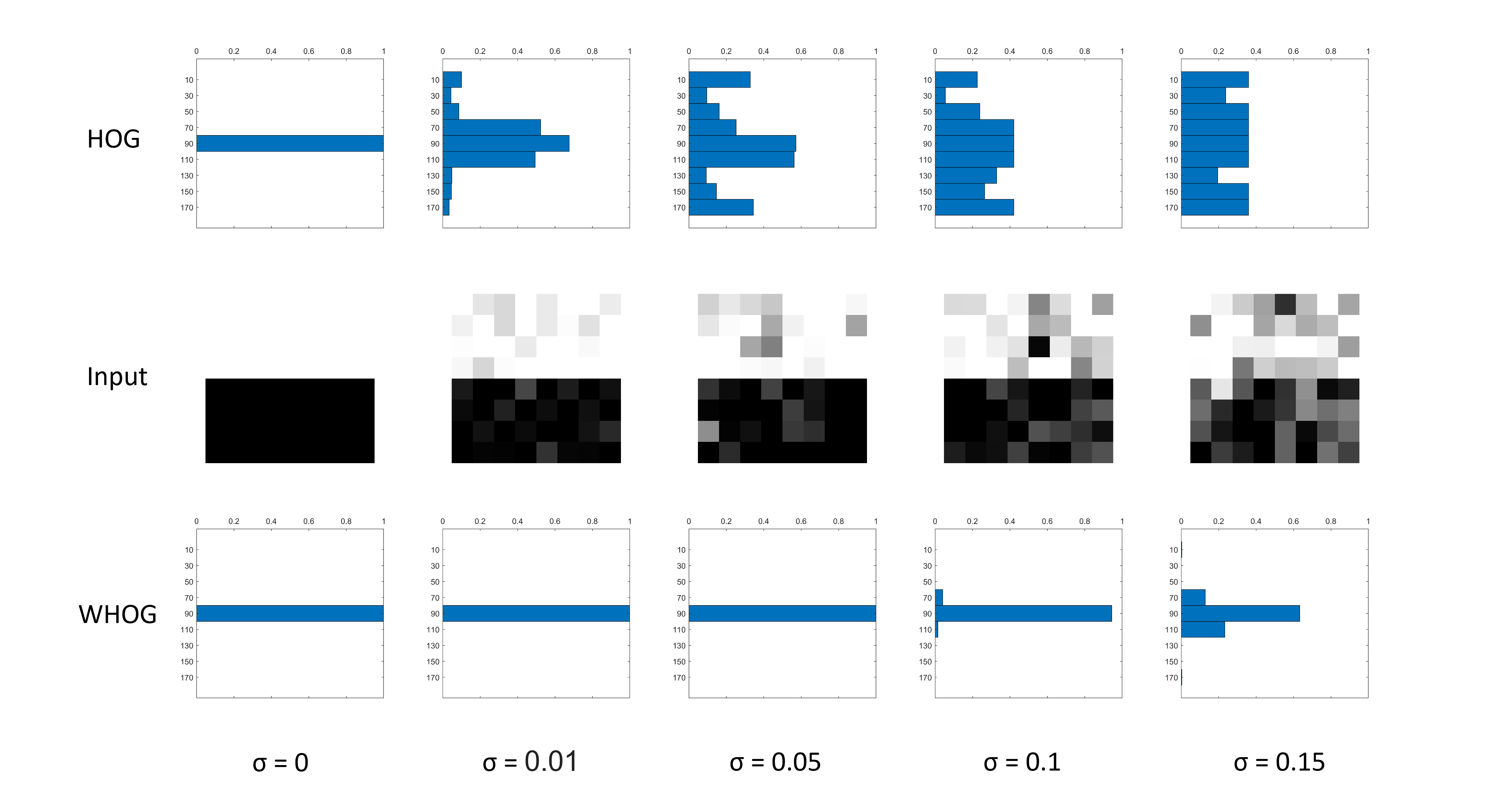}
  \includegraphics[width=0.48\linewidth]{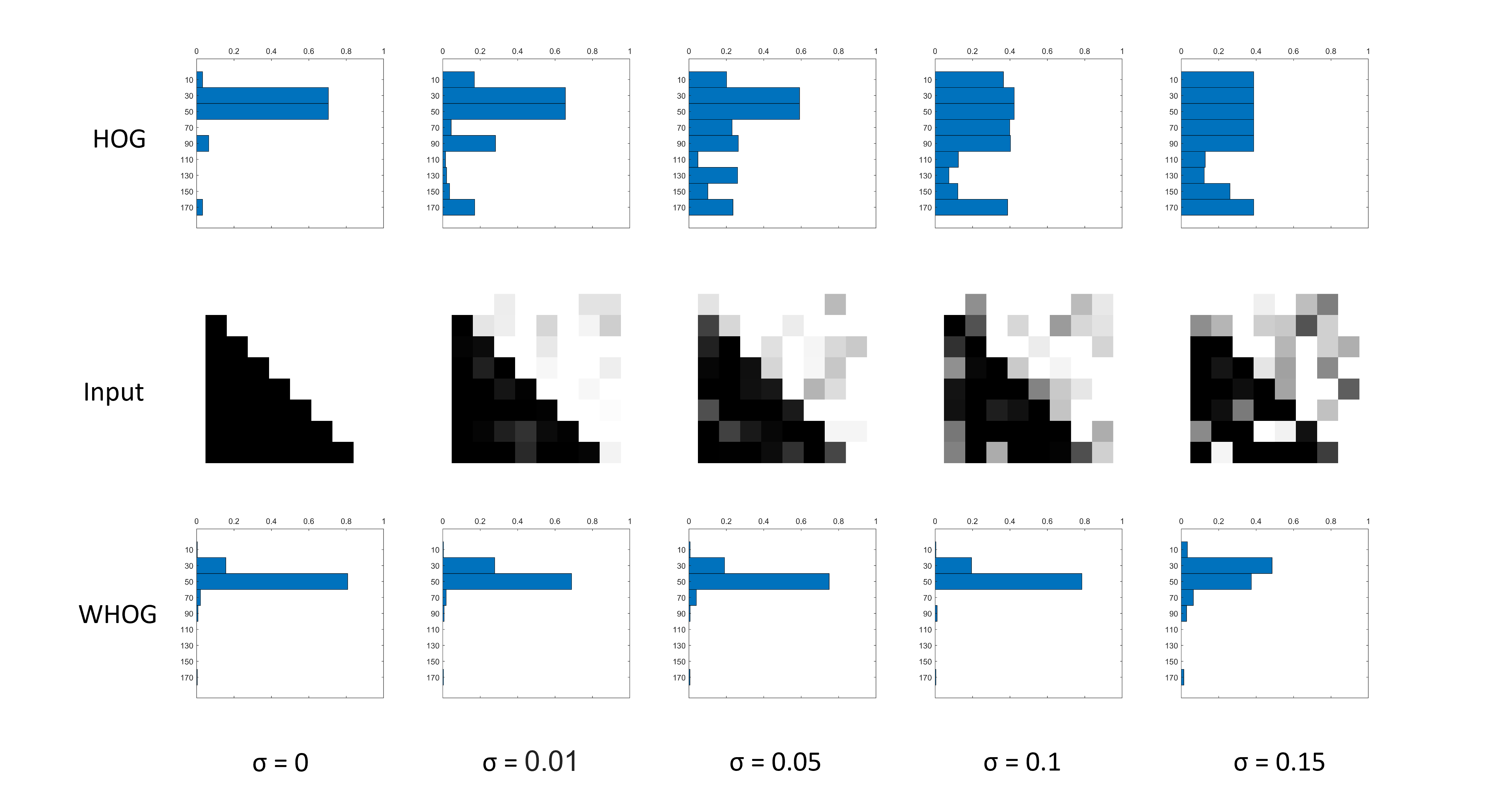}
  \caption{HOG vs WHOG under different noise level: for both methods, we use 8 by 8 patch and 9 bins.}\label{compare}
\end{figure}

Notice that the way we get directionality (\ref{theta}) is very different from HOG (\ref{theta_hog}). The direction of HOG is based on a gradient, which is sensitive to noise. WHOG, on the other hand, considers all the sub-transport routes and the direction is determined by grid points, making WHOG more robust. The examples in Fig.\ref{compare} show that both methods work well when there is no noise. However, when more noise is added (Gaussian noise with mean=0, $\sigma=\{0,0.01,0.01,0.1,0.15\}$), WHOG almost doesn't change while HOG gradually lose directionality until the histogram is evenly distributed to all the bins.

\begin{figure}[H]
  \centering
  \includegraphics[width=0.2\linewidth]{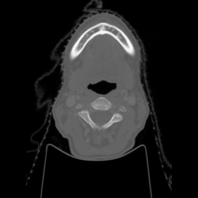}
  \includegraphics[width=0.2\linewidth]{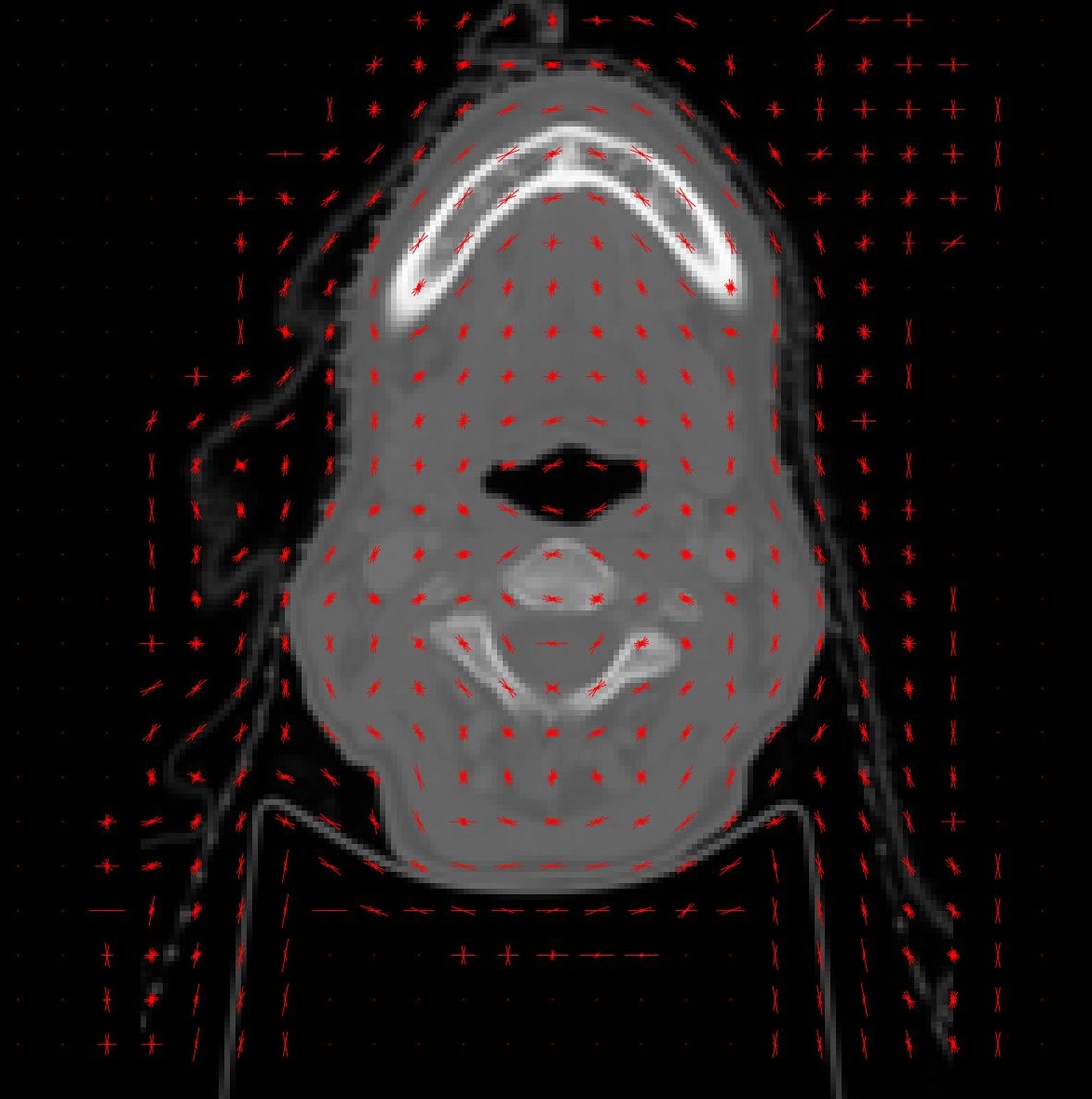}
  \includegraphics[width=0.2\linewidth]{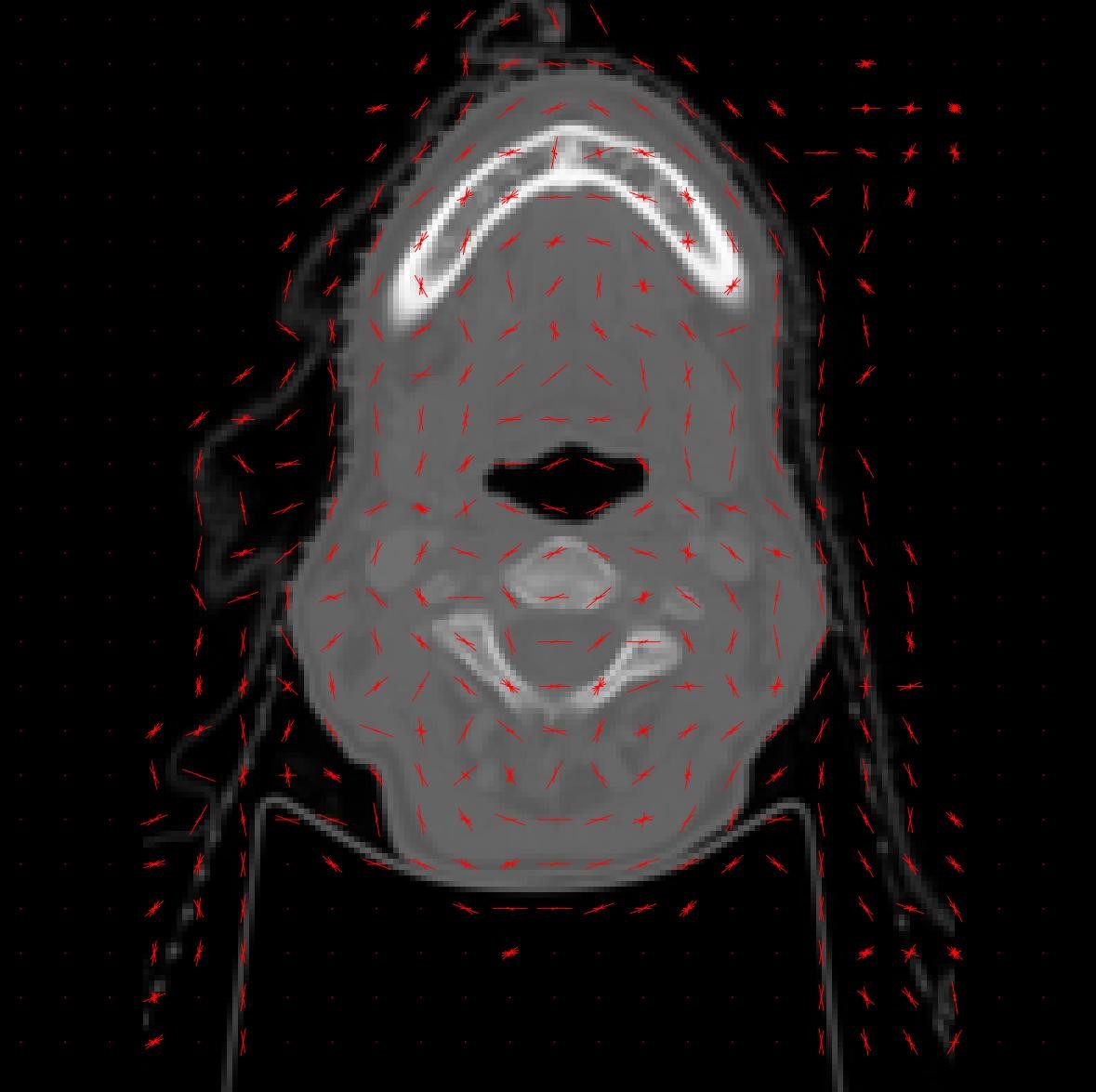}
  \caption{An example of a head and neck slice from PDDCA dataset. Left: original slice, Middle: Rose plot from HOG, Right: Rose plot from WHOG.}\label{c2}
\end{figure}

Bellow is an example of a medical image (Fig.~\ref{c2}). The rose plots illustrate the local orientations extracted by two methods. We observe that WHOG extracts the shapes of soft tissues, and we find that our proposed feature detector is better at extracting clear directionality, especially in the tissue area.

\subsection{Brain Glioma}
Biological evidence indicates that some of the flows within tumor regions are connected tumor severity. Those tumors with trajectories which are more spread out, seem to have greater disorganization and are malignant.

We examined brain MRI images of glioblastoma multiforme patients from TCIA. WHOG local directionality feature vectors were computed within the tumor regions for each sample. By randomly taking the same amount of patches for different patients (tumor sizes differ), we computed the entropy of those distributions of directionality. Interestingly, we found significant survival differences (log-rank test p-value $=0.008$ vs the p-value of HOG $=0.34$) between the high entropy group and low entropy group (cut at the median value of entropy). The high entropy group which has more flows in many directions is more malignant, while the low entropy group with more steady flows are more benign. 

\begin{figure}[H]
  \centering
  \includegraphics[width=0.45\linewidth]{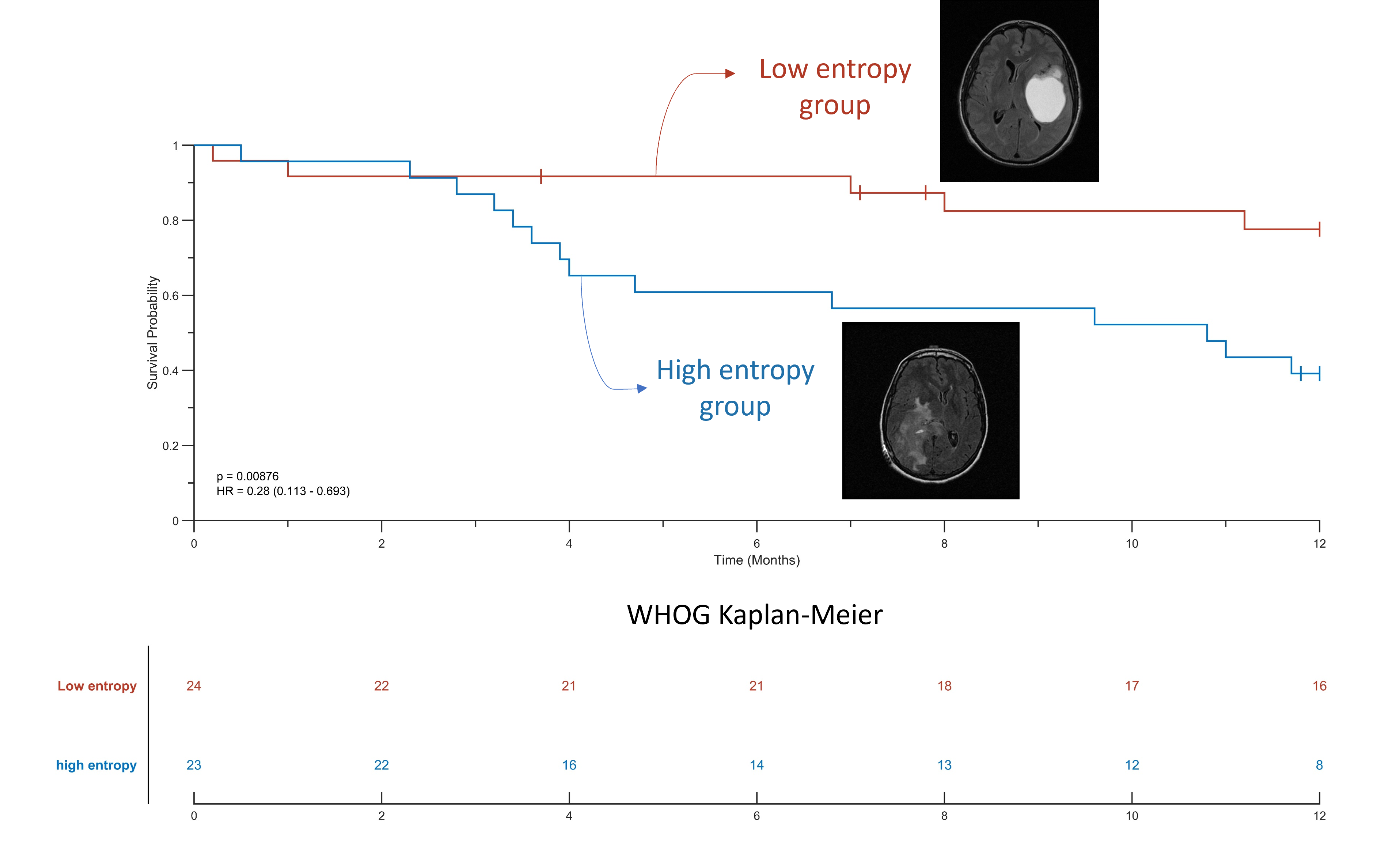}
  \includegraphics[width=0.45\linewidth]{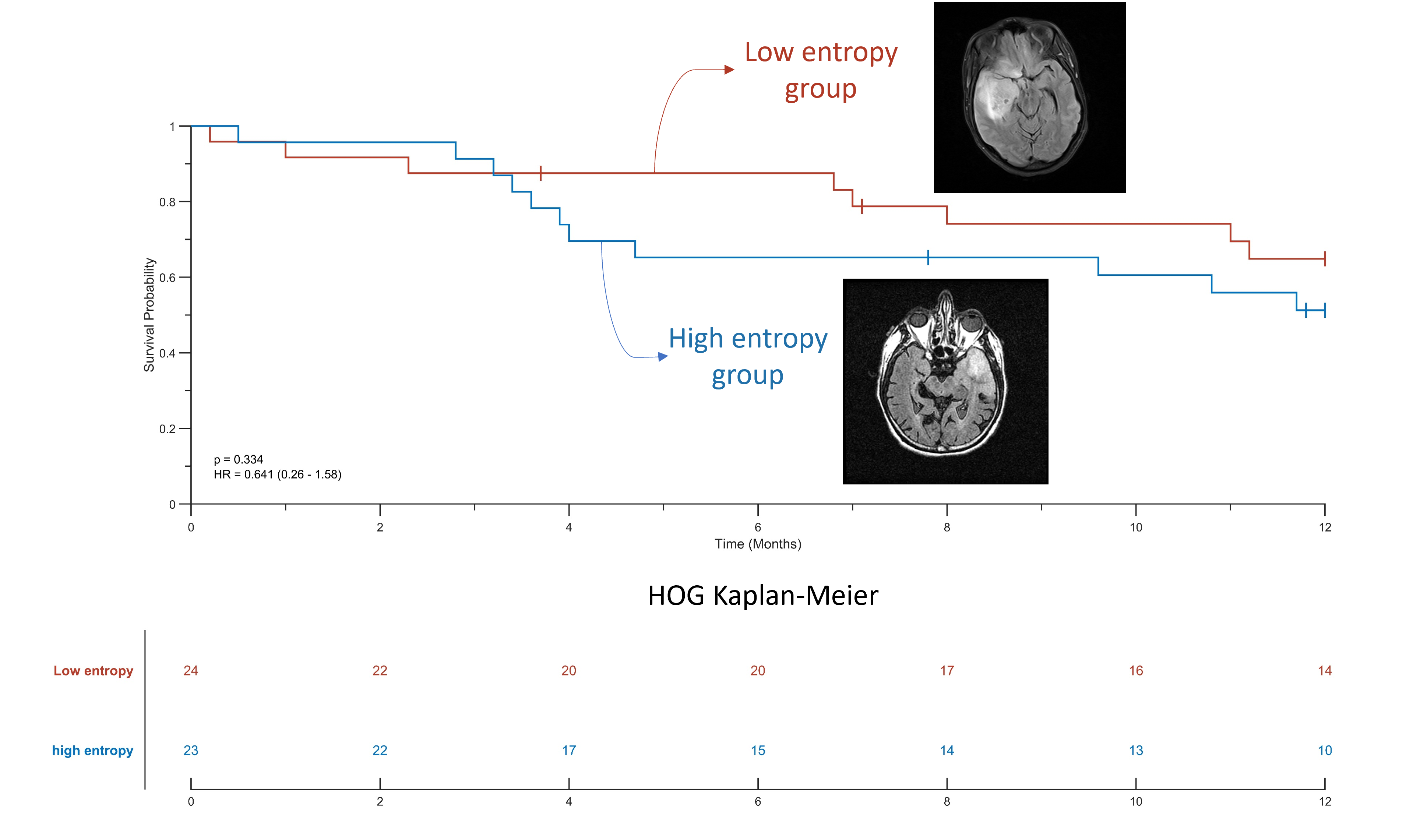}
  \caption{Kaplan-Meier plots of survival difference between the high entropy group and the low entropy group each with a characterized slice from a patient in each group from WHOG (left) and HOG (right).}\label{gb}
\end{figure}

We compared a few slices from the largest and lowest entropy patients for both WHOG and HOG methods. We find that some patients are in the high entropy group from the HOG method is due to noise. We think the reason that the WHOG gets significant survival difference is the result of its unique ability to handle noise which helps to extract truly biological information of the tumor local fluid flows. 

To test whether our WHOG feature is related to scanner manufacturers and the magnet strengths. We computed the Wilcoxon rank sum test between GE vs non-GE; 1.5T vs 3T. p-values are: 0.6, 0.8 respectively. Thus for these data, the WHOG feature does not seem to correlate with GE or the MRI strength.

\subsection{Smoothing and Edge Detection}
We now illustrate our method via a smoothing and edge detection example.
\begin{figure}[H]
  \centering
  \includegraphics[width=0.25\linewidth]{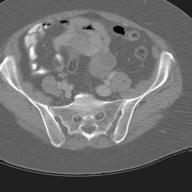}
  \includegraphics[width=0.25\linewidth]{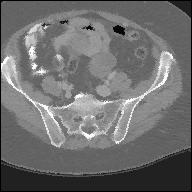}
  \includegraphics[width=0.25\linewidth]{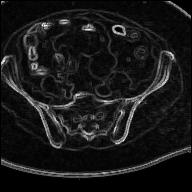}
  \caption{An example of optimal transport based smoothing and edge detection. Left: original image. Middle: smoothed image. Right: detected edges.}\label{ddd}
\end{figure}

Referring to Figure~\ref{ddd}, we see that the smoothed image reduces much of the background noise while all the edges are preserved. Most of the edges are extracted with good quality.

\section{Discussion and Conclusion}
The present work makes use of optimal transport in image local analysis to extract information such as directionality and edge information. Because of its robustness, we find our method performs well even in the presence of noise. As a general local analysis model, we considered the optimal transport between local patches and 1-color patches or 2-color patches to get interpretable results, which is useful in multiple classical image processing tasks. 

We got similar conclusion about image directionality and tumor heterogeneity in \cite{prasanna_co-occurrence_2016}. They did classifications based on their HOG based feature vetors. In principle, we can replace their HOG feature by our WHOG feature. We hope by doing this, classification accuracy can be further improved.

We plan to further evaluate ability of our method to study pathology images. The microscopic flows in these high resolution images contain information for which our proposed method may better capture directionality, and thus distinguish different biological processes that are going on in different tissues.

The method can be easily modified to 3D case by looking at the transport map between a local cube to a pure color cube. We also want to test on the 3D extension so that the directionality is within the transverse plane, frontal, and sagittal directions.

Generally speaking, we can consider the transport between a local patch and any other distribution on that patch. Notice that the process of computing the optimal transport map acts like a local nonlinear filter, but unlike commonly used linear filters which may reduce the information in the original data, all the information is preserved, and may be recovered by projection. By considering the transport map, the results are in the product space with the pairwise information. This may be incorporated in a neural network as an upsampling layer. 


%
%
%
\bibliographystyle{splncs04}
\bibliography{refs}

\begin{thebibliography}{10}
\providecommand{\url}[1]{\texttt{#1}}
\providecommand{\urlprefix}{URL }
\providecommand{\doi}[1]{https://doi.org/#1}

\bibitem{Arjovsky2017}
Arjovsky, M., Chintala, S., Bottou, L.: {Wasserstein GAN}. arxiv.org
  \textbf{1701.07875} (2017)

\bibitem{CarMaa14}
Carlen, E.A., Maas, J.: An analog of the 2-{W}asserstein metric in
  non-commutative probability under which the {F}ermionic {F}okker--{P}lanck
  equation is gradient flow for the entropy. Communications in Mathematical
  Physics  \textbf{331}(3),  887--926 (2014)

\bibitem{romt}
Chen, X., Tran, A.P., Elkin, R., Benveniste, H., Tannenbaum, A.: Visualizing
  fluid flows via regularized optimal mass transport with applications to
  neuroscience  (01 2022)

\bibitem{QIQQA-UKDEU}
Dalal, N., Triggs, B.: Histograms of oriented gradients for human detection (6
  2005). \doi{10.1109/CVPR.2005.177}

\bibitem{Evans1999}
Evans, L.C., Gangbo, W.: {Differential equations methods for the
  Monge-Kantorovich mass transfer problem}. Memoirs of the American
  Mathematical Society  \textbf{137}(653) (1999). \doi{10.1090/memo/0653}

\bibitem{93808}
Freeman, W., Adelson, E.: The design and use of steerable filters. IEEE
  Transactions on Pattern Analysis and Machine Intelligence  \textbf{13}(9),
  891--906 (1991). \doi{10.1109/34.93808}

\bibitem{Haker2004}
Haker, S., Zhu, L., Tannenbaum, A., Angenent, S.: {Optimal mass transport for
  registration and warping}. International Journal of Computer Vision
  \textbf{60}(3),  225--240 (2004). \doi{10.1023/B:VISI.0000036836.66311.97}

\bibitem{LK}
Kantorovich, L.V.: On a problem of monge. CR (Doklady) Acad. Sci. URSS (NS)
  \textbf{3},  225--226 (1948)

\bibitem{Lowe2004}
Lowe, D.G.: Distinctive image features from scale-invariant keypoints.
  International Journal of Computer Vision  \textbf{60}(2),  91--110 (Nov
  2004). \doi{10.1023/b:visi.0000029664.99615.94},
  \url{https://doi.org/10.1023/b:visi.0000029664.99615.94}

\bibitem{Statement2020}
Mathews, J.C., Nadeem, S., Pouryahya, M., Belkhatir, Z., Deasy, J.O., Levine,
  A.J., Tannenbaum, A.R.: {Functional network analysis reveals an immune
  tolerance mechanism in cancer}. Proceedings of the National Academy of
  Sciences  \textbf{117}(28),  16339--16345 (jul 2020).
  \doi{10.1073/pnas.2002179117},
  \url{http://www.pnas.org/lookup/doi/10.1073/pnas.2002179117}

\bibitem{MitMie16}
Mittnenzweig, M., Mielke, A.: An entropic gradient structure for lindblad
  equations and coupling of quantum systems to macroscopic models. J. Stat.
  Physics  \textbf{167}(2) (2017)

\bibitem{9266088}
Pouryahya, M., Oh, J., Javanmard, P., Mathews, J.C., Belkhatir, Z., Deasy,
  J.O., Tannenbaum, A.: awcluster: A novel integrative network-based clustering
  of multiomics for subtype analysis of cancer data. IEEE/ACM Transactions on
  Computational Biology and Bioinformatics (01), ~1--1 (nov 5555).
  \doi{10.1109/TCBB.2020.3039511}

\bibitem{prasanna_co-occurrence_2016}
Prasanna, P., Tiwari, P., Madabhushi, A.: Co-occurrence of {Local}
  {Anisotropic} {Gradient} {Orientations} ({CoLlAGe}): {A} new radiomics
  descriptor. Scientific Reports  \textbf{6}(1),  37241 (Dec 2016).
  \doi{10.1038/srep37241}, \url{http://www.nature.com/articles/srep37241}

\bibitem{Rachev1998}
Rachev, S.T., R{\"{u}}schendorf, L.: {Mass Transportation Problems: Volume I:
  Theory}. Probability and its Applications, Springer, Berlin (1998).
  \doi{10.1007/b98894}, \url{http://link.springer.com/10.1007/b98894}

\bibitem{rr}
Rachev, S.T., R{\"u}schendorf, L.: Mass Transportation Problems: {V}olumes {I}
  and {II}. Springer Science \& Business Media (1998)

\bibitem{villani1}
Villani, C.: Topics in Optimal Transportation. American Mathematical Soc.
  (2003)

\bibitem{villani2}
Villani, C.: Optimal Transport: Old and New, vol.~338. Springer Science \&
  Business Media (2008)

\bibitem{Zhu2021.06.17.448878}
Zhu, J., Oh, J.H., Deasy, J.O., Tannenbaum, A.: vwcluster: A network based
  clustering of multi-omics breast cancer data based on vector-valued optimal
  transport. bioRxiv  (2021). \doi{10.1101/2021.06.17.448878},
  \url{https://www.biorxiv.org/content/early/2021/06/18/2021.06.17.448878}

\bibitem{aapm}
Zhu, J.;~Thor, M.A.A.O.J.R.A.D.J.T.A.: A non-parametric analysis to identify
  high-risk vs. low-risk anatomic regions associated with reduced overall
  survival in the rtog 0617 clinical trial data  (06 2021)

\end{thebibliography}
\end{document}